\documentclass[a4paper,fleqn]{cas-dc}
\usepackage{natbib}
\usepackage{amsmath}

\def\tsc#1{\csdef{#1}{\textsc{\lowercase{#1}}\xspace}}
\tsc{WGM}
\tsc{QE}
\tsc{EP}
\tsc{PMS}
\tsc{BEC}
\tsc{DE}

\begin{document}
\let\WriteBookmarks\relax
\def\floatpagepagefraction{1}
\def\textpagefraction{.001}
\shorttitle{}
\shortauthors{J}

\title [mode = title]{SUNRISE-3D: Sharp UNveiling of AGN feedback Regulation and its Impact on Star-formation at the cosmic noon Epoch \tnotemark[1]} \tnotetext[1]{This paper is based on a presentation on the SHARP science case for the Extremely Large Telescope.}

\author[1]{G. Vietri}[type=editor,
                        orcid=0000-0001-9155-8875]
\cormark[1]
\ead{giustina.vietri@inaf.it}

\credit{Writing - original draft, Conceptualization, Methodology}

\affiliation[1]{organization={INAF – Istituto di Astrofisica Spaziale e Fisica Cosmica Milano},
                addressline={Via Alfonso Corti 12}, 
                city={Milano},
                postcode={20133}, 
                country={Italy}}

\author[1]{S. Bisogni}
\credit{Investigation, Writing - review \& editing}

\author[2]{E. Piconcelli}
\credit{Investigation, Writing - review \& editing}

\author[3]{F. Ricci}
\credit{Investigation, Writing - review \& editing}

\author[1]{P. Franzetti}
\credit{Software - ETC developer}

\author[1]{A. Gargiulo}
\credit{Investigation, Writing - review \& editing}

\author[1]{C. Mancini}
\credit{Investigation, Writing - review \& editing}

\affiliation[2]{organization={INAF - Osservatorio Astronomico di Roma},
                addressline={Via Frascati 33, Monte Porzio Catone}, 
                city={Rome},
                postcode={00040}, 
                country={Italy}}

\affiliation[3]{organization={Department of Mathematics and Physics, Roma Tre University},
                addressline={Via della Vasca Navale 84}, 
                city={Rome},
                postcode={00146}, 
                country={Italy}}

\cortext[cor1]{Corresponding author}

\begin{abstract}
To better understand the role of AGN-driven outflows as a mechanism for heating or sweeping up gas over distances comparable to the size of the galaxy in its evolution, and to explore their physical characteristics as a function of AGN and host galaxy properties, it is necessary to have a statistical sample of AGNs selected from a uniform sample of galaxies with spectroscopic coverage of key rest-frame optical emission lines.
To assess the impact of AGN-driven outflows on their host galaxies, we need to derive the mass and energy carried by the outflows, as well as correlations of these quantities with both AGN and host galaxy properties, in order to reveal their effects on the galaxy population and constrain the physical mechanisms driving the outflows.
The availability of adaptive-optics-assisted 3D spectroscopy with the ELT multi-IFU instrument SHARP/VESPER enables the construction of spatially resolved outflow property maps, providing instantaneous outflow rates across the entire field of view without assuming outflow geometry, and thus significantly reducing the uncertainties compared to methods based on longslit spectroscopy.
Furthermore, combining these maps with resolved star formation rate (SFR) maps allows a direct comparison between outflow properties and star-formation activity across the galaxy, providing key insights into how AGN feedback regulates star formation down to sub-kpc scales.
By applying this approach to a representative sample of galaxies at cosmic noon ($1.5 < z < 2.5$), spanning a wide range of stellar masses from low-mass systems ($M_\star = 10^{8-10}\,M_\odot$) to the massive end ($M_\star > 10^{10}\,M_\odot$), we aim to systematically investigate the interplay between AGN activity, outflows, and star formation in the galaxy population as a whole.
\end{abstract}

 \begin{keywords}
AGN - feedback - cosmic noon - outflows - IFU
\end{keywords}

\maketitle

\section{Scientific rationale}

Theoretical models of galaxy formation predict more stars than are actually observed: only $\sim$20\% of the baryonic content within dark matter halos forms stellar mass (e.g.~\citealt{Guo2010}), with the efficiency peaking in halos of $\sim10^{12}$\,M$_\odot$,  corresponding to a stellar mass of $\sim10^{10.5}$\,M$_\odot$, and declining at both lower and higher masses. This implies that some mechanism must actively regulate star formation (SF) across the full mass spectrum.
At the low-mass end, supernova explosions and winds from massive stars can drive outflows that redistribute gas and metals, the so-called stellar feedback (e.g.~\citealt{Hopkins2014}), as the shallower gravitational potential makes the interstellar medium (ISM) more susceptible to energy injection. In more massive galaxies, however, the energy released by stars is insufficient to counteract the deep gravitational potential, and feedback from accreting supermassive black holes (SMBHs) becomes the dominant mechanism. Active galactic nuclei (AGN) inject energy both radiatively and kinetically into their host galaxies (e.g.~\citealt{Peng2010}; \citealt{DiMatteo2005}; \citealt{Pillepich2018}), regulating star formation and explaining observed correlations between black hole mass and galaxy properties (see~\citealt{Fabian2012} for a review). Despite significant observational progress over the past decade, from focused studies of the most luminous quasars to large statistical surveys of more typical galaxies (see Sect.~\ref{sec:AGN}), focusing on a single gas phase yields only a partial view of galactic outflows. A blind, spatially resolved, multi-phase census of feedback across the entire galaxy population is still missing, and the cosmic noon ($1.5 \lesssim z \lesssim 2.5$), marking the peak of both cosmic star formation and black hole accretion, represents the ideal epoch to conduct it. The Adaptive-Optics (AO)-corrected multi-IFU SHARP/VESPER, covering 1.2--2.4\,$\mu$m at R\,$\sim$\,3000 and resolving structures down to 30\,mas ($\approx$0.3\,kpc at $z\sim2$), will simultaneously unveil the ionized and (likely) neutral gas components across the full galaxy  population, making a spatially resolved, multi-phase census of this kind possible for the first time.

\subsection{The role of the AGN feedback}\label{sec:AGN}

The multi-phase nature of galactic outflows is now well established. The ionized component is probed via rest-frame optical lines such as [OIII], which traces the narrow-line region (NLR) gas (\citealt{Kakkad2020}), and rest-frame UV lines such as CIV, which traces winds on nuclear scales (\citealt{Vietri2020, Vietri2025}); the neutral phase through blueshifted absorption features such as NaID (\citealt{Rupke2011}; \citealt{Davies2024}); and the cold molecular gas via asymmetric profiles of transitions such as CO (\citealt{Cicone2014}).

Powerful outflows are predominantly observed in luminous AGN and QSOs, reaching velocities of several thousand km s$^{-1}$ (e.g. \citealt{Cano-Diaz2012}; \citealt{ShenHo2014}; \citealt{VillarMartin2020}). For example, in the WISE/SDSS-selected hyper-luminous quasar (WISSH) survey, 85 targets with $L_{\mathrm{bol}} > 10^{47}$ erg s$^{-1}$ (\citealt{Saccheo2023}) at z=1.5-4.5, the ionized component, traced for 18 QSOs by targeting the [O III] line with NIR LBT/LUCI spectrograph, shows exceptional luminosities and velocities (up to several thousand km s$^{-1}$), mass-outflow rates of several thousand M$_\odot$ yr$^{-1}$ and kinetic powers amounting to a few percent of the bolometric luminosity ($L_{\rm Bol}$), demonstrating highly efficient winds that can sweep out ionized gas (\citealt{Bischetti2017}, \citealt{Vietri2018}).

By analyzing heterogeneous, large sample of AGN, \citet{Fiore2017} showed that galaxies hosting powerful AGN winds have molecular gas depletion times and gas fractions that are 3–10 times lower than those galaxies with similar star formation rates, stellar masses, and redshifts. This implies that, at high AGN bolometric luminosities, a substantial fraction of the molecular gas may be disrupted by the outflow, shifting much of it into atomic or ionized phases, though the impact 
appears more limited at lower stellar masses.

\subsubsection{Statistical approach}

Statistical studies of ionized outflows at cosmic noon became possible with the advent of near-infrared multi-object spectrographs as MOSFIRE, KMOS, and FMOS enabling observations of 500--1000 galaxies at a time and revealing the incidence and properties of galactic winds across the 
population.

In the MOSFIRE/MOSDEF survey, a sample of 159 AGNs at 1.4 $\le$ z $\le$ 3.8 with $L\rm_{Bol}$=10$^{44-47}$ erg/s (identified via X-ray, IR, and/or optical wavelengths) was embedded in a uniformly selected galaxy sample of both star‐forming and quiescent systems spanning $\sim$3 orders of magnitude in stellar mass (\citealt{Leung2019}). Outflows were identified via H$\beta$, [O III], H$\alpha$ and [N II] emission: detected in 17\% of the AGNs, seven times more frequently than in a mass-matched sample of inactive galaxies. The outflows are fast ($\sim$400-3500 km/s), and from 2D spectra the physical extent along the slit direction was measured to be $\sim$0.3–11 kpc. The incidence of outflows among AGNs was found to be independent of host-galaxy stellar mass, and outflows were found in both star-forming and quiescent galaxies. However, the absence of spatially resolved measurements introduces uncertainties, as outflows are almost certainly geometrically complex (e.g., \citealt{Kakkad2020}).

A complementary view comes from \citealt{FS2019}, who presented a large sample (152) of AGNs at z$\approx$1–3, drawn from a sample of 599 galaxies observed with integral field spectroscopy (IFS) with KMOS in seeing‐limited mode covering H$\alpha$, [N II], and [S II]. While the IFS data provided 2D measurements of the outflow size (reaching a spatial resolution of 4 kpc at z$\sim$2), the [O III] and H$\beta$ emission lines are not covered. The [O III] line is a particularly valuable tracer of NLR gas and ionised outflows because it lies far from other bright emission lines, and is less prone to contamination than [N II]. They found that $\sim$15\% of galaxies show AGN outflows, with the incidence, strength and velocity strongly correlated with stellar mass, and independent of star‐formation activity.
A dedicated AGN survey, the KMOS AGN Survey at High redshift (KASHz; \citealt{Harrison2016}) has provided no-AO spatially resolved observations for hundreds of X-ray selected AGN, revealing ionised gas velocities that are likely indicative of outflows in about 50\% of the examined sample. However, again seeing-limited observations are not sufficient to trace outflows and their properties in detail.

A significant advance in angular resolution came with the SUPER programme (\citealt{Circosta2018}), which used  VLT/SINFONI in AO-assisted mode to achieve 2--4\,kpc resolution at $z\sim2$. Ionized outflows traced by [OIII] were found to be ubiquitous across the full sample of 33 X-ray selected AGN (L$_\mathrm{bol} = 10^{44- 47}$ erg s$^{-1}$; 21 Type 1 and 12 Type 2)
selected without any bias toward systems known to host strong winds, with spatially resolved emission in 35\%  of Type~1 and 86\% of Type~2 sources extending over 2--10\,kpc (\citealt{Kakkad2020}; \citealt{Tozzi2024}). 
Type 2 AGN host faster ionized outflows than their Type 1 counterparts within the same bolometric luminosity range. The velocities of ionized outflows detected in the SUPER sample are comparable to the escape speeds of their dark matter halos and are generally high enough to reach distances of 30–50 kpc from the galaxy center. These outflows can therefore expel gas from the baryonic disk and/or provide preventive-mode feedback by injecting energy and driving turbulence in the surrounding medium. In either case, they have the potential to reduce or even quench star formation.

In the low-mass regime (M$_\star < 10^{10}$ M$_\odot$), AGN have been detected in local galaxies (e.g., \citealt{Sartori2015}; \citealt{Penny2018}; \citealt{Reines2020}; \citealt{Mezcua2023}), but probing their outflow properties at higher redshifts has been challenging due to the limited sensitivity of ground-based near-infrared instruments. The launch of JWST has started to change this picture: surveys like JADES have observed [O III] and H$\alpha$ emission from dwarf galaxies (52 objects) at $z \sim 3.5$--8.5 (M$_\star$ $\sim 10^{7-9}$ M$_\odot$) using 1D multi-object spectroscopy (\citealt{Carniani2024}). Emission lines were detected with S/N $>5$ (here S/N is defined as the ratio of the flux peak of the line and the sensitivity level of the observation at the same wavelength), and outflows were identified in roughly 25\% of the sample, significantly higher than in local low-mass galaxies. These results open a new window for studying feedback in low-mass galaxies at early cosmic times, a regime that was previously inaccessible from the ground.

Recent JWST 1D spectra (R $\sim 1000$) of individual quiescent galaxies at $z > 2$ have revealed weak ionized gas emission lines clearly linked to AGN activity, which would have been undetectable with ground-based observations (\citealt{Belli2024}; \citealt{Carnall2023}; \citealt{DEugenio2024}). Thanks to the exceptional sensitivity of JWST/NIRSpec, the Blue Jay survey (\citealt{Belli2024}) found that low-luminosity AGN are remarkably common in high-redshift quiescent galaxies (\citealt{Bugiani2025}).  In this survey, 113 galaxies at $1.7 < z < 3.5$ were analyzed to study the demographics and properties of neutral gas outflows traced by NaID absorption, covering both star-forming and passive galaxies with stellar masses $M_\star > 10^{9.5}$ M$_\odot$. The results show that interstellar NaID absorption is widespread in massive galaxies ($M_\star > 10^{10}$ M$_\odot$) at $z \sim 2$, and that these neutral outflows are likely AGN-driven. This implies that powerful AGN-driven neutral gas outflows are common among massive galaxies at this epoch and may constitute a significant mechanism for rapid quenching (\citealt{Davies2024}).

\subsubsection{Overcoming current limitations: a path forward.} 

Despite the progress outlined above, most statistical studies of galactic outflows have so far suffered from restricted spectral coverage, often limited to H$\alpha$ alone, which is challenging to model due to its proximity to [N II] and possible contamination from broad-line region emission. Longslit/1D spectra further provide only limited information on the ionized gas, severely limiting robust constraints on the geometry and true spatial extent of the outflows. Moreover, most studies have focused on a single gas phase, providing an incomplete picture of the total outflow impact. The AO-corrected multi-IFU SHARP/VESPER, covering 1.2--2.4\,$\mu$m at R\,$\approx$\,3000, overcomes these limitations within a single setup: the simultaneous coverage of [O III]+H$\beta$, H$\alpha$+[N II]+[S II], and Na I D enables the study of feedback mechanisms through ionized and neutral outflows at $1.5<z<2.5$ in a statistical sample of both low- and high-mass galaxies. By simultaneously tracing star formation via H$\alpha$ and the neutral and ionized gas phases through the Na I D and [O III] lines, respectively, we will be able to characterize outflows in unprecedented detail. The VESPER spaxel size (31\,mas\,$\approx$\,300\,pc at $z\approx2$) will resolve star-forming clumps and outflow substructures with unmatched precision, enabling robust conclusions on the global impact of AGN-driven feedback on the galaxy population.

\begin{figure*}
    \centering
    \includegraphics[width=0.7\textwidth]{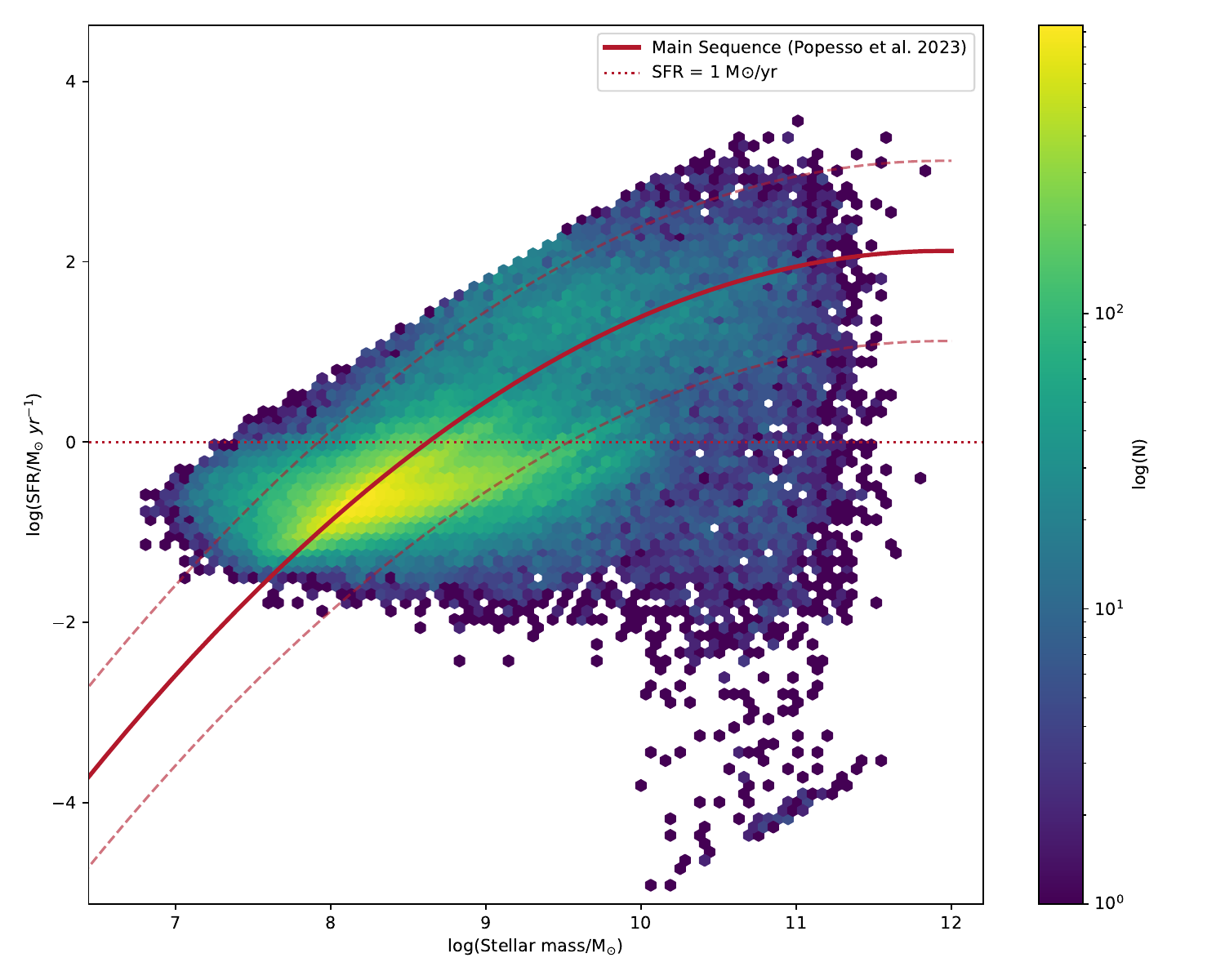}
   \caption{Star formation versus stellar mass for COSMOS-Web galaxies at 1.5 $<$ z $<$ 2.5 (\citealt{Shuntov2025}). The star-forming main sequence at z$\sim$2 from \cite{Popesso2023} is also shown as solid red curve, with a dispersion of $\pm$ 1 dex. The adopted observational SFR limit of 1 M$_\odot$/yr is indicated by the red dotted horizontal line..}\label{fig:sfr_mstar}
\end{figure*}

\section{Objectives}

The multi-IFU capability of VESPER (N=12), combined with the collecting power of the ELT, will provide the S/N required to detect even the faintest galactic winds in a statistical sample of galaxies spanning the full mass range, from low-mass systems ($M_\star = 10^{8-10}$\,M$_\odot$) to the massive end ($M_\star > 10^{10}$\,M$_\odot$). This will allow us to spatially resolve both neutral and ionized gas and to achieve the following goals:

\noindent \textbullet\ \textbf{Obtaining a spatially resolved view of the ionized and neutral outflows.} For the ionized phase, spaxel-by-spaxel analysis of the [O III] line profile will yield the centroid, velocity shift, width, and luminosity of both systemic and broad/blueshifted components, and outflow distance, allowing us to reconstruct the outflow kinematics and geometry. The ionizing source will be identified through the BPT diagnostic diagram ([O III]/H$\beta$ vs. [N II]/H$\alpha$; \citealt{Baldwin1981}), complemented by multi-band photometric AGN identification (see Sect.~\ref{sec:sample}), separating AGN-driven from stellar ionization. Electron densities will be primarily estimated from the spatially resolved [S II]$\lambda\lambda$6716,6731 doublet ratio, sensitive over the range $n_{\rm e} \approx 10^2$--$10^{3.5}$ cm$^{-3}$, though this diagnostic is known to underestimate the true outflow density in AGN-photoionized gas (\citealt{Davies2020}). We will therefore complement it with the ionization-parameter method of \citet{BaronN2019}, which derives $n_{\rm e}$ from the [O III]/H$\beta$ and [N II]/H$\alpha$ line ratios combined with the spatially resolved outflow distance, all quantities accessible within a single SHARP/VESPER setup. Together, these measurements will allow us to derive key physical quantities such as the mass outflow rate and kinetic power, thereby constraining the efficiency of the winds in influencing their host galaxies. By constructing spaxel-by-spaxel mass outflow rate maps with spatially variable density and velocity, we will derive instantaneous outflow rates across the full field of view without assuming a global outflow geometry, avoiding the large biases that affect single-aperture or longslit-based estimates (\citealt{Kakkad2022}). For the neutral component, the relative weakness of the NaID absorption feature requires either analysis of the integrated spectrum or Voronoi binning of the 3D data cube to enhance the S/N (see Sect. \ref{sec:ETC} for further details).

\noindent \textbullet\ \textbf{Obtaining a spatially resolved view of star formation} over $\sim$10\,Myr timescales, comparable to those of the outflow. By disentangling SF-associated H$\alpha$ emission from AGN contributions, we will construct SFR maps and compare them directly with the ionized and neutral outflow maps, searching for spatial signatures of negative and positive AGN feedback. A spatial anti-correlation between outflow and SFR on a spaxel-by-spaxel basis, i.e. the suppression of star formation precisely where the outflow is detected, constitutes direct evidence of negative feedback (\citealt{Cresci2015}), while positive feedback may manifest as an enhancement at the outflow edges where gas compression triggers new star formation (\citealt{Cresci2015}; \citealt{CresciMaiolino2018}). We acknowledge that H$\alpha$ traces SF on $\sim$10\,Myr timescales while AGN activity varies on much shorter timescales; any causal interpretation will therefore be discussed in the context of the relevant timescale uncertainties.

\noindent \textbullet\ \textbf{Estimating the incidence of ionized and neutral outflows as a function of host galaxy properties.} By probing the ionized phase through [OIII] and the neutral phase through NaID, we will establish how outflow occurrence depends on stellar mass and SFR, and whether it is causally linked to ongoing star-forming activity as traced by H$\alpha$. A delayed-feedback scenario, in which outflows preferentially occur in high-SFR systems at fixed stellar mass (\citealt{Woo2017}), will be directly testable with our dataset.

\noindent \textbullet\ \textbf{Investigating the scaling relations.} Simulations predict that AGN feedback correlates more strongly with black hole mass than with instantaneous bolometric luminosity, pointing to a dependence on integrated rather than current AGN activity (\citealt{Piotrowska2022}). Testing this requires measuring outflow properties (e.g., velocity, distance) against $L_{\rm Bol}$, estimated from spectral energy distribution (SED) fitting or bolometric corrections to the continuum at 5100\,\AA, and black hole mass, derived directly for Type~1 AGN via Balmer lines (\citealt{Shen2013}) or indirectly for Type~2 AGN (\citealt{Baron2019}).

\subsection{The SUNRISE sample}\label{sec:sample}

Fields with deep multi-wavelength coverage are the natural choice for a programme of this kind. The COSMOS field, with its extensive photometric data from X-ray to radio (\citealt{Marchesi2016}; \citealt{Weaver2022}), provides the ideal foundation: its source density of $\sim$200000 objects over 2\,deg$^2$ at $1.5 < z < 2.5$ implies $\sim$10--15 galaxies per VESPER/SHARP pointing of 24$\times$70\,arcsec$^2$, making it perfectly matched to the multi-IFU capability of VESPER. From this dataset, stellar masses and integrated SFRs can be derived through SED fitting, while AGN can be identified via X-ray, MIR/radio, or optical selection, all without pre-selecting galaxies in a particular evolutionary stage or with unusual properties. Using MCAO-assisted observations, we will achieve angular resolutions of $\sim$30\,mas for high-SFR galaxies, while for low-SFR systems the data quality will be comparable to NIRSpec@JWST ($\sim$100\,mas).

\subsubsection{ETC calculation}\label{sec:ETC}

Our goal is to investigate 3D outflow and star-formation properties, extending down to low-mass ($M_\star < 10^{10}\,M_\odot$) systems and up to the passive population at the high-mass end. We aim to take advantage of the flux limits achievable with the ELT/SHARP–VESPER instrument. 
From Fig. \ref{fig:sfr_mstar}, which shows the distribution of COSMOS-Web galaxies in the instantaneous SFR–stellar mass plane (\citealt{Shuntov2025}), a SFR limit of $\sim 1~M_\odot~\mathrm{yr}^{-1}$ is indicated, representing the threshold above which galaxies can be followed up within a reasonable observing time. Indeed, by converting the expected SFR into H$\alpha$ emission line fluxes, we estimate that a SFR~$\sim 1$~$M_\odot$~yr$^{-1}$ corresponds to an H$\alpha$ flux of $\sim 7 \times 10^{-18}$~erg~s$^{-1}$~cm$^{-2}$ (assuming a Chabrier IMF) at $z=2$ (\citealt{Kennicutt1998}).
We used the SHARP ETC (v0.5) to simulate a galaxy with $M_\star$=10$^{9}$ $M_{\odot}$ at $z\sim2.35$ with $f_\mathrm{H\alpha} \sim 7\times10^{-18}$~erg~s$^{-1}$~cm$^{-2}$, FWHM = 500 km/s ($\sim 37$ \AA) at $\lambda = 22000$~\AA, and a Sersic radius of $R_\mathrm{e} = 250$~mas (\citealt{Shuntov2025}). An integration time of $\sim 20$ hours per source is required to achieve S/N~$\sim 6$ (after binning 7 spectral pixels, corresponding to 20 \AA$\sim$270 km/s) for the H$\alpha$ line at its peak in the central pixel, and S/N~$\sim 1.5$ at 0.3 arcsec, assuming a spatial resampling of $3\times3$ pixels (100~mas). 
These exposure times should also permit spatially resolved [O III] outflow measurements. The feasibility of detecting the NaID absorption feature instead depends critically on the S/N achieved on the continuum rather than on emission line fluxes. The NaID doublet is an intrinsically weak absorption feature, and its detection at $z \sim 2$ has so far been demonstrated only in integrated 1D spectra of massive galaxies ($M_\star \gtrsim 10^{10}\,M_\odot$) using deep JWST/NIRSpec observations 
(\citealt{Davies2024, Bugiani2025}). We therefore expect the NaID analysis to be feasible for the high-mass end of our sample, where the continuum  S/N will be sufficient for individual detections or modest Voronoi binning of the SHARP/VESPER data cubes. For lower-mass systems ($M_\star < 10^{10}\,M_\odot$), we will rely on the integrated spectrum of each galaxy, or on spectral stacking across sub-samples, to place statistical constraints on the neutral gas outflow properties.

For brighter targets in terms of SFR or outflows, higher S/N can be reached with such exposure times by adopting the native angular resolution of SHARP ($\sim$30 mas), allowing us to probe sub-kpc scales.

\section{Comparison with other facilities}

The ELT will host several first- and second-generation instruments designed to probe galaxy evolution at high redshift. For the specific goal of conducting a statistical, spatially resolved study of AGN-driven feedback at cosmic noon, a combination of multiplexing, angular resolution, and broad near-infrared spectral coverage is required. In this context, VESPER provides instrumental capabilities that are well matched to the scientific objectives of this programme.

Compared to ELT/HARMONI, which provides single-IFU integral-field spectroscopy over a contiguous field of view, VESPER enables the simultaneous IFU observation of up to 12 galaxies within a single pointing, making it particularly well suited for building statistically significant samples of galaxies and AGN at  z$\sim$2 within realistic observing times.

The wavelength coverage of VESPER (1.2--2.4~$\mu$m) is optimally matched to the key rest-frame optical diagnostics at $1.5 < z < 2.5$, including [OIII], H$\beta$, H$\alpha$, [NII], [SII], and NaID. Access to these emission and absorption lines within a single instrumental setup enables a self-consistent, multi-phase characterization of ionized and neutral gas.

Relative to JWST/NIRSpec, VESPER provides a factor of $\sim$3--4 improvement in angular resolution thanks to the ELT aperture and adaptive optics correction, reaching spatial scales of $\sim$30~mas ($\approx$300~pc at $z\sim2$). This resolution is critical for resolving the spatial structure and kinematics of galactic outflows, which are otherwise blurred together at the angular resolution of JWST.

\section{Conclusions}

AGN-driven outflows are widely invoked as a key mechanism regulating star formation and driving the co-evolution of galaxies and supermassive black holes, yet their global impact and physical coupling to the host galaxy remain poorly constrained, particularly at the peak epoch of galaxy assembly.
Current observational limitations have restricted most studies to small samples, single gas phases, or seeing-limited spatial resolutions, preventing a comprehensive and statistically robust assessment of feedback processes.

The SUNRISE programme, enabled by the multi-IFU, AO-assisted capabilities of ELT/VESPER, will overcome these limitations by providing the first blind, spatially resolved, multi-phase census of AGN-driven outflows in a representative galaxy population at cosmic noon.
By simultaneously mapping ionized and neutral gas outflows together with star formation on sub-kpc scales, this survey will allow us to directly test how and where AGN feedback operates, whether it suppresses or potentially triggers star formation, and how its efficiency depends on galaxy mass and AGN properties.

\section*{Acknowledgments}
The SHARP team acknowledges support by Bando Ricerca Fondamentale INAF 2022, Techno-Grant "SHARP" - 1.05.12.02.01 and Bando Ricerca Fondamentale INAF 2024, Large-Grant "SHARP" - 1.05.24.01.01.\\

\printcredits

\bibliographystyle{cas-model2-names}

\bibliography{bib_revised}

@ARTICLE{Bischetti2017,
       author = {{Bischetti}, M. and {Piconcelli}, E. and {Vietri}, G. and {Bongiorno}, A. and {Fiore}, F. and {Sani}, E. and {Marconi}, A. and {Duras}, F. and {Zappacosta}, L. and {Brusa}, M. and {Comastri}, A. and {Cresci}, G. and {Feruglio}, C. and {Giallongo}, E. and {La Franca}, F. and {Mainieri}, V. and {Mannucci}, F. and {Martocchia}, S. and {Ricci}, F. and {Schneider}, R. and {Testa}, V. and {Vignali}, C.},
        title = "{The WISSH quasars project. I. Powerful ionised outflows in hyper-luminous quasars}",
      journal = {\aap},
         year = 2017,
        month = feb,
       volume = {598},
          eid = {A122},
        pages = {A122},
          doi = {10.1051/0004-6361/201629301},
archivePrefix = {arXiv},
       eprint = {1612.03728},
 primaryClass = {astro-ph.GA},
       adsurl = {https://ui.adsabs.harvard.edu/abs/2017A&A...598A.122B}
}

@ARTICLE{Bugiani2025,
       author = {{Bugiani}, Letizia and {Belli}, Sirio and {Park}, Minjung and {Davies}, Rebecca L. and {Mendel}, J. Trevor and {Johnson}, Benjamin D. and {Khoram}, Amir H. and {Benton}, Chlo{\"e} and {Cimatti}, Andrea and {Conroy}, Charlie and {Emami}, Razieh and {Leja}, Joel and {Li}, Yijia and {Maheson}, Gabriel and {Mathews}, Elijah P. and {Naidu}, Rohan P. and {Nelson}, Erica J. and {Tacchella}, Sandro and {Terrazas}, Bryan A. and {Weinberger}, Rainer},
        title = "{Active Galactic Nucleus Feedback in Quiescent Galaxies at Cosmic Noon Traced by Ionized Gas Emission}",
      journal = {\apj},
         year = 2025,
        month = mar,
       volume = {981},
       number = {1},
          eid = {25},
        pages = {25},
          doi = {10.3847/1538-4357/adaeaf},
archivePrefix = {arXiv},
       eprint = {2406.08547},
 primaryClass = {astro-ph.GA},
       adsurl = {https://ui.adsabs.harvard.edu/abs/2025ApJ...981...25B}
}

@ARTICLE{VillarMartin2020,
       author = {{Villar Mart{\'\i}n}, M. and {Perna}, M. and {Humphrey}, A. and {Castro Rodr{\'\i}guez}, N. and {Binette}, L. and {P{\'e}rez Gonz{\'a}lez}, P.~G. and {Mateos}, S. and {Cabrera Lavers}, A.},
        title = "{Peculiar emission line spectra of core extremely red BOSS quasars at z{\ensuremath{\sim}}2-3: orientation and/or evolution?}",
      journal = {\aap},
         year = 2020,
        month = feb,
       volume = {634},
          eid = {A116},
        pages = {A116},
          doi = {10.1051/0004-6361/201937086},
archivePrefix = {arXiv},
       eprint = {2001.03379},
 primaryClass = {astro-ph.GA},
       adsurl = {https://ui.adsabs.harvard.edu/abs/2020A&A...634A.116V}
}

@ARTICLE{ShenHo2014,
       author = {{Shen}, Yue and {Ho}, Luis C.},
        title = "{The diversity of quasars unified by accretion and orientation}",
      journal = {\nat},
         year = 2014,
        month = sep,
       volume = {513},
       number = {7517},
        pages = {210-213},
          doi = {10.1038/nature13712},
archivePrefix = {arXiv},
       eprint = {1409.2887},
 primaryClass = {astro-ph.GA},
       adsurl = {https://ui.adsabs.harvard.edu/abs/2014Natur.513..210S}
}

@ARTICLE{Cano-Diaz2012,
       author = {{Cano-D{\'\i}az}, M. and {Maiolino}, R. and {Marconi}, A. and {Netzer}, H. and {Shemmer}, O. and {Cresci}, G.},
        title = "{Observational evidence of quasar feedback quenching star formation at high redshift}",
      journal = {\aap},
         year = 2012,
        month = jan,
       volume = {537},
          eid = {L8},
        pages = {L8},
          doi = {10.1051/0004-6361/201118358},
archivePrefix = {arXiv},
       eprint = {1112.3071},
 primaryClass = {astro-ph.CO},
       adsurl = {https://ui.adsabs.harvard.edu/abs/2012A&A...537L...8C}
}

@ARTICLE{Carnall2023,
       author = {{Carnall}, Adam C. and {McLure}, Ross J. and {Dunlop}, James S. and {McLeod}, Derek J. and {Wild}, Vivienne and {Cullen}, Fergus and {Magee}, Dan and {Begley}, Ryan and {Cimatti}, Andrea and {Donnan}, Callum T. and {Hamadouche}, Massissilia L. and {Jewell}, Sophie M. and {Walker}, Sam},
        title = "{A massive quiescent galaxy at redshift 4.658}",
      journal = {\nat},
         year = 2023,
        month = jul,
       volume = {619},
       number = {7971},
        pages = {716-719},
          doi = {10.1038/s41586-023-06158-6},
archivePrefix = {arXiv},
       eprint = {2301.11413},
 primaryClass = {astro-ph.GA},
       adsurl = {https://ui.adsabs.harvard.edu/abs/2023Natur.619..716C}
}

@ARTICLE{Carniani2024,
       author = {{Carniani}, Stefano and {Venturi}, Giacomo and {Parlanti}, Eleonora and {de Graaff}, Anna and {Maiolino}, Roberto and {Arribas}, Santiago and {Bonaventura}, Nina and {Boyett}, Kristan and {Bunker}, Andrew J. and {Cameron}, Alex J. and {Charlot}, Stephane and {Chevallard}, Jacopo and {Curti}, Mirko and {Curtis-Lake}, Emma and {Eisenstein}, Daniel J. and {Giardino}, Giovanna and {Hausen}, Ryan and {Kumari}, Nimisha and {Maseda}, Michael V. and {Nelson}, Erica and {Perna}, Michele and {Rix}, Hans-Walter and {Robertson}, Brant and {Del Pino}, Bruno Rodr{\'\i}guez and {Sandles}, Lester and {Scholtz}, Jan and {Simmonds}, Charlotte and {Smit}, Renske and {Tacchella}, Sandro and {{\"U}bler}, Hannah and {Williams}, Christina C. and {Willott}, Chris and {Witstok}, Joris},
        title = "{JADES: The incidence rate and properties of galactic outflows in low-mass galaxies across 3 < z < 9}",
      journal = {\aap},
         year = 2024,
        month = may,
       volume = {685},
          eid = {A99},
        pages = {A99},
          doi = {10.1051/0004-6361/202347230},
archivePrefix = {arXiv},
       eprint = {2306.11801},
 primaryClass = {astro-ph.GA},
       adsurl = {https://ui.adsabs.harvard.edu/abs/2024A&A...685A..99C}
}

@article{Cresci2015,
  author    = {Cresci, G. and Mainieri, V. and Brusa, M. and others},
  title     = {},
  journal   = {ApJ},
  volume    = {799},
  pages     = {82},
  year      = {2015}
}

@ARTICLE{Circosta2018,
       author = {{Circosta}, C. and {Mainieri}, V. and {Padovani}, P. and {Lanzuisi}, G. and {Salvato}, M. and {Harrison}, C.~M. and {Kakkad}, D. and {Puglisi}, A. and {Vietri}, G. and {Zamorani}, G. and {Cicone}, C. and {Husemann}, B. and {Vignali}, C. and {Balmaverde}, B. and {Bischetti}, M. and {Bongiorno}, A. and {Brusa}, M. and {Carniani}, S. and {Civano}, F. and {Comastri}, A. and {Cresci}, G. and {Feruglio}, C. and {Fiore}, F. and {Fotopoulou}, S. and {Karim}, A. and {Lamastra}, A. and {Magnelli}, B. and {Mannucci}, F. and {Marconi}, A. and {Merloni}, A. and {Netzer}, H. and {Perna}, M. and {Piconcelli}, E. and {Rodighiero}, G. and {Schinnerer}, E. and {Schramm}, M. and {Schulze}, A. and {Silverman}, J. and {Zappacosta}, L.},
        title = "{SUPER. I. Toward an unbiased study of ionized outflows in z {\ensuremath{\sim}} 2 active galactic nuclei: survey overview and sample characterization}",
      journal = {\aap},
         year = 2018,
        month = nov,
       volume = {620},
          eid = {A82},
        pages = {A82},
          doi = {10.1051/0004-6361/201833520},
archivePrefix = {arXiv},
       eprint = {1809.04858},
 primaryClass = {astro-ph.GA},
       adsurl = {https://ui.adsabs.harvard.edu/abs/2018A&A...620A..82C}
}

@ARTICLE{DiMatteo2005,
       author = {{Di Matteo}, Tiziana and {Springel}, Volker and {Hernquist}, Lars},
        title = "{Energy input from quasars regulates the growth and activity of black holes and their host galaxies}",
      journal = {\nat},
         year = 2005,
        month = feb,
       volume = {433},
       number = {7026},
        pages = {604-607},
          doi = {10.1038/nature03335},
archivePrefix = {arXiv},
       eprint = {astro-ph/0502199},
 primaryClass = {astro-ph},
       adsurl = {https://ui.adsabs.harvard.edu/abs/2005Natur.433..604D}
}

@ARTICLE{Fabian2012,
       author = {{Fabian}, A.~C.},
        title = "{Observational Evidence of Active Galactic Nuclei Feedback}",
      journal = {\araa},
         year = 2012,
        month = sep,
       volume = {50},
        pages = {455-489},
          doi = {10.1146/annurev-astro-081811-125521},
archivePrefix = {arXiv},
       eprint = {1204.4114},
 primaryClass = {astro-ph.CO},
       adsurl = {https://ui.adsabs.harvard.edu/abs/2012ARA&A..50..455F}
}

@ARTICLE{Fiore2017,
       author = {{Fiore}, F. and {Feruglio}, C. and {Shankar}, F. and {Bischetti}, M. and {Bongiorno}, A. and {Brusa}, M. and {Carniani}, S. and {Cicone}, C. and {Duras}, F. and {Lamastra}, A. and {Mainieri}, V. and {Marconi}, A. and {Menci}, N. and {Maiolino}, R. and {Piconcelli}, E. and {Vietri}, G. and {Zappacosta}, L.},
        title = "{AGN wind scaling relations and the co-evolution of black holes and galaxies}",
      journal = {\aap},
         year = 2017,
        month = may,
       volume = {601},
          eid = {A143},
        pages = {A143},
          doi = {10.1051/0004-6361/201629478},
archivePrefix = {arXiv},
       eprint = {1702.04507},
 primaryClass = {astro-ph.GA},
       adsurl = {https://ui.adsabs.harvard.edu/abs/2017A&A...601A.143F}
}

@ARTICLE{FS2019,
       author = {{F{\"o}rster Schreiber}, N.~M. and {{\"U}bler}, H. and {Davies}, R.~L. and {Genzel}, R. and {Wisnioski}, E. and {Belli}, S. and {Shimizu}, T. and {Lutz}, D. and {Fossati}, M. and {Herrera-Camus}, R. and {Mendel}, J.~T. and {Tacconi}, L.~J. and {Wilman}, D. and {Beifiori}, A. and {Brammer}, G.~B. and {Burkert}, A. and {Carollo}, C.~M. and {Davies}, R.~I. and {Eisenhauer}, F. and {Fabricius}, M. and {Lilly}, S.~J. and {Momcheva}, I. and {Naab}, T. and {Nelson}, E.~J. and {Price}, S.~H. and {Renzini}, A. and {Saglia}, R. and {Sternberg}, A. and {van Dokkum}, P. and {Wuyts}, S.},
        title = "{The KMOS$^{3D}$ Survey: Demographics and Properties of Galactic Outflows at z = 0.6-2.7}",
      journal = {\apj},
         year = 2019,
        month = apr,
       volume = {875},
       number = {1},
          eid = {21},
        pages = {21},
          doi = {10.3847/1538-4357/ab0ca2},
archivePrefix = {arXiv},
       eprint = {1807.04738},
 primaryClass = {astro-ph.GA},
       adsurl = {https://ui.adsabs.harvard.edu/abs/2019ApJ...875...21F}
}

@ARTICLE{Guo2010,
       author = {{Guo}, Qi and {White}, Simon and {Li}, Cheng and {Boylan-Kolchin}, Michael},
        title = "{How do galaxies populate dark matter haloes?}",
      journal = {\mnras},
         year = 2010,
        month = may,
       volume = {404},
       number = {3},
        pages = {1111-1120},
          doi = {10.1111/j.1365-2966.2010.16341.x},
archivePrefix = {arXiv},
       eprint = {0909.4305},
 primaryClass = {astro-ph.CO},
       adsurl = {https://ui.adsabs.harvard.edu/abs/2010MNRAS.404.1111G}
}

@ARTICLE{Hopkins2014,
       author = {{Hopkins}, Philip F. and {Kere{\v{s}}}, Du{\v{s}}an and {O{\~n}orbe}, Jos{\'e} and {Faucher-Gigu{\`e}re}, Claude-Andr{\'e} and {Quataert}, Eliot and {Murray}, Norman and {Bullock}, James S.},
        title = "{Galaxies on FIRE (Feedback In Realistic Environments): stellar feedback explains cosmologically inefficient star formation}",
      journal = {\mnras},
         year = 2014,
        month = nov,
       volume = {445},
       number = {1},
        pages = {581-603},
          doi = {10.1093/mnras/stu1738},
archivePrefix = {arXiv},
       eprint = {1311.2073},
 primaryClass = {astro-ph.CO},
       adsurl = {https://ui.adsabs.harvard.edu/abs/2014MNRAS.445..581H}
}

@ARTICLE{Leung2019,
       author = {{Leung}, Gene C.~K. and {Coil}, Alison L. and {Aird}, James and {Azadi}, Mojegan and {Kriek}, Mariska and {Mobasher}, Bahram and {Reddy}, Naveen and {Shapley}, Alice and {Siana}, Brian and {Fetherolf}, Tara and {Fornasini}, Francesca M. and {Freeman}, William R. and {Price}, Sedona H. and {Sanders}, Ryan L. and {Shivaei}, Irene and {Zick}, Tom},
        title = "{The MOSDEF Survey: A Census of AGN-driven Ionized Outflows at z = 1.4-3.8}",
      journal = {\apj},
         year = 2019,
        month = nov,
       volume = {886},
       number = {1},
          eid = {11},
        pages = {11},
          doi = {10.3847/1538-4357/ab4a7c},
archivePrefix = {arXiv},
       eprint = {1905.13338},
 primaryClass = {astro-ph.GA},
       adsurl = {https://ui.adsabs.harvard.edu/abs/2019ApJ...886...11L}
}

@article{Mezcua2023,
  author    = {Mezcua, M. and others},
  title     = {},
  journal   = {MNRAS},
  volume    = {518},
  pages     = {3245},
  year      = {2023}
}

@article{Penny2018,
  author    = {Penny, S. J. and Masters, K. L. and Smethurst, R. J. and others},
  title     = {},
  journal   = {MNRAS},
  volume    = {476},
  pages     = {979},
  year      = {2018}
}

@ARTICLE{Peng2010,
       author = {{Peng}, Ying-jie and {Lilly}, Simon J. and {Kova{\v{c}}}, Katarina and {Bolzonella}, Micol and {Pozzetti}, Lucia and {Renzini}, Alvio and {Zamorani}, Gianni and {Ilbert}, Olivier and {Knobel}, Christian and {Iovino}, Angela and {Maier}, Christian and {Cucciati}, Olga and {Tasca}, Lidia and {Carollo}, C. Marcella and {Silverman}, John and {Kampczyk}, Pawel and {de Ravel}, Loic and {Sanders}, David and {Scoville}, Nicholas and {Contini}, Thierry and {Mainieri}, Vincenzo and {Scodeggio}, Marco and {Kneib}, Jean-Paul and {Le F{\`e}vre}, Olivier and {Bardelli}, Sandro and {Bongiorno}, Angela and {Caputi}, Karina and {Coppa}, Graziano and {de la Torre}, Sylvain and {Franzetti}, Paolo and {Garilli}, Bianca and {Lamareille}, Fabrice and {Le Borgne}, Jean-Francois and {Le Brun}, Vincent and {Mignoli}, Marco and {Perez Montero}, Enrique and {Pello}, Roser and {Ricciardelli}, Elena and {Tanaka}, Masayuki and {Tresse}, Laurence and {Vergani}, Daniela and {Welikala}, Niraj and {Zucca}, Elena and {Oesch}, Pascal and {Abbas}, Ummi and {Barnes}, Luke and {Bordoloi}, Rongmon and {Bottini}, Dario and {Cappi}, Alberto and {Cassata}, Paolo and {Cimatti}, Andrea and {Fumana}, Marco and {Hasinger}, Gunther and {Koekemoer}, Anton and {Leauthaud}, Alexei and {Maccagni}, Dario and {Marinoni}, Christian and {McCracken}, Henry and {Memeo}, Pierdomenico and {Meneux}, Baptiste and {Nair}, Preethi and {Porciani}, Cristiano and {Presotto}, Valentina and {Scaramella}, Roberto},
        title = "{Mass and Environment as Drivers of Galaxy Evolution in SDSS and zCOSMOS and the Origin of the Schechter Function}",
      journal = {\apj},
         year = 2010,
        month = sep,
       volume = {721},
       number = {1},
        pages = {193-221},
          doi = {10.1088/0004-637X/721/1/193},
archivePrefix = {arXiv},
       eprint = {1003.4747},
 primaryClass = {astro-ph.CO},
       adsurl = {https://ui.adsabs.harvard.edu/abs/2010ApJ...721..193P}
}

@ARTICLE{Pillepich2018,
       author = {{Pillepich}, Annalisa and {Springel}, Volker and {Nelson}, Dylan and {Genel}, Shy and {Naiman}, Jill and {Pakmor}, R{\"u}diger and {Hernquist}, Lars and {Torrey}, Paul and {Vogelsberger}, Mark and {Weinberger}, Rainer and {Marinacci}, Federico},
        title = "{Simulating galaxy formation with the IllustrisTNG model}",
      journal = {\mnras},
         year = 2018,
        month = jan,
       volume = {473},
       number = {3},
        pages = {4077-4106},
          doi = {10.1093/mnras/stx2656},
archivePrefix = {arXiv},
       eprint = {1703.02970},
 primaryClass = {astro-ph.GA},
       adsurl = {https://ui.adsabs.harvard.edu/abs/2018MNRAS.473.4077P}
}

@article{Reines2020,
  author    = {Reines, A. E. and others},
  title     = {},
  journal   = {ApJ},
  volume    = {888},
  pages     = {36},
  year      = {2020}
}

@ARTICLE{Saccheo2023,
       author = {{Saccheo}, I. and {Bongiorno}, A. and {Piconcelli}, E. and {Testa}, V. and {Bischetti}, M. and {Bisogni}, S. and {Bruni}, G. and {Cresci}, G. and {Feruglio}, C. and {Fiore}, F. and {Grazian}, A. and {Luminari}, A. and {Lusso}, E. and {Mainieri}, V. and {Maiolino}, R. and {Marconi}, A. and {Ricci}, F. and {Tombesi}, F. and {Travascio}, A. and {Vietri}, G. and {Vignali}, C. and {Zappacosta}, L. and {La Franca}, F.},
        title = "{The WISSH quasars project. XI. The mean spectral energy distribution and bolometric corrections of the most luminous quasars}",
      journal = {\aap},
         year = 2023,
        month = mar,
       volume = {671},
          eid = {A34},
        pages = {A34},
          doi = {10.1051/0004-6361/202244296},
archivePrefix = {arXiv},
       eprint = {2211.07677},
 primaryClass = {astro-ph.GA},
       adsurl = {https://ui.adsabs.harvard.edu/abs/2023A&A...671A..34S}
}

@article{Sartori2015,
  author    = {Sartori, L. F. and Schawinski, K. and Treister, E. and others},
  title     = {},
  journal   = {MNRAS},
  volume    = {454},
  pages     = {3722},
  year      = {2015}
}

@ARTICLE{Vietri2018,
       author = {{Vietri}, G. and {Piconcelli}, E. and {Bischetti}, M. and {Duras}, F. and {Martocchia}, S. and {Bongiorno}, A. and {Marconi}, A. and {Zappacosta}, L. and {Bisogni}, S. and {Bruni}, G. and {Brusa}, M. and {Comastri}, A. and {Cresci}, G. and {Feruglio}, C. and {Giallongo}, E. and {La Franca}, F. and {Mainieri}, V. and {Mannucci}, F. and {Ricci}, F. and {Sani}, E. and {Testa}, V. and {Tombesi}, F. and {Vignali}, C. and {Fiore}, F.},
        title = "{The WISSH quasars project. IV. Broad line region versus kiloparsec-scale winds}",
      journal = {\aap},
         year = 2018,
        month = sep,
       volume = {617},
          eid = {A81},
        pages = {A81},
          doi = {10.1051/0004-6361/201732335},
archivePrefix = {arXiv},
       eprint = {1802.03423},
 primaryClass = {astro-ph.GA},
       adsurl = {https://ui.adsabs.harvard.edu/abs/2018A&A...617A..81V}
}

@ARTICLE{Belli2024,
       author = {{Belli}, Sirio and {Park}, Minjung and {Davies}, Rebecca L. and {Mendel}, J. Trevor and {Johnson}, Benjamin D. and {Conroy}, Charlie and {Benton}, Chlo{\"e} and {Bugiani}, Letizia and {Emami}, Razieh and {Leja}, Joel and {Li}, Yijia and {Maheson}, Gabriel and {Mathews}, Elijah P. and {Naidu}, Rohan P. and {Nelson}, Erica J. and {Tacchella}, Sandro and {Terrazas}, Bryan A. and {Weinberger}, Rainer},
        title = "{Star formation shut down by multiphase gas outflow in a galaxy at a redshift of 2.45}",
      journal = {\nat},
         year = 2024,
        month = jun,
       volume = {630},
       number = {8015},
        pages = {54-58},
          doi = {10.1038/s41586-024-07412-1},
archivePrefix = {arXiv},
       eprint = {2308.05795},
 primaryClass = {astro-ph.GA},
       adsurl = {https://ui.adsabs.harvard.edu/abs/2024Natur.630...54B}
}

@ARTICLE{DEugenio2024,
       author = {{D'Eugenio}, Francesco and {P{\'e}rez-Gonz{\'a}lez}, Pablo G. and {Maiolino}, Roberto and {Scholtz}, Jan and {Perna}, Michele and {Circosta}, Chiara and {{\"U}bler}, Hannah and {Arribas}, Santiago and {B{\"o}ker}, Torsten and {Bunker}, Andrew J. and {Carniani}, Stefano and {Charlot}, Stephane and {Chevallard}, Jacopo and {Cresci}, Giovanni and {Curtis-Lake}, Emma and {Jones}, Gareth C. and {Kumari}, Nimisha and {Lamperti}, Isabella and {Looser}, Tobias J. and {Parlanti}, Eleonora and {Rix}, Hans-Walter and {Robertson}, Brant and {Rodr{\'\i}guez Del Pino}, Bruno and {Tacchella}, Sandro and {Venturi}, Giacomo and {Willott}, Chris J.},
        title = "{A fast-rotator post-starburst galaxy quenched by supermassive black-hole feedback at z = 3}",
      journal = {Nature Astronomy},
         year = 2024,
        month = nov,
       volume = {8},
        pages = {1443-1456},
          doi = {10.1038/s41550-024-02345-1},
archivePrefix = {arXiv},
       eprint = {2308.06317},
 primaryClass = {astro-ph.GA},
       adsurl = {https://ui.adsabs.harvard.edu/abs/2024NatAs...8.1443D}
}

@ARTICLE{Tozzi2024,
       author = {{Tozzi}, G. and {Cresci}, G. and {Perna}, M. and {Mainieri}, V. and {Mannucci}, F. and {Marconi}, A. and {Kakkad}, D. and {Marasco}, A. and {Brusa}, M. and {Bertola}, E. and {Bischetti}, M. and {Carniani}, S. and {Cicone}, C. and {Circosta}, C. and {Fiore}, F. and {Feruglio}, C. and {Harrison}, C.~M. and {Lamperti}, I. and {Netzer}, H. and {Piconcelli}, E. and {Puglisi}, A. and {Scholtz}, J. and {Vietri}, G. and {Vignali}, C. and {Zamorani}, G.},
        title = "{SUPER: VIII. Fast and furious at z {\ensuremath{\sim}} 2: Obscured type-2 active nuclei host faster ionised winds than type-1 systems}",
      journal = {\aap},
         year = 2024,
        month = oct,
       volume = {690},
          eid = {A141},
        pages = {A141},
          doi = {10.1051/0004-6361/202450162},
archivePrefix = {arXiv},
       eprint = {2407.04099},
 primaryClass = {astro-ph.GA},
       adsurl = {https://ui.adsabs.harvard.edu/abs/2024A&A...690A.141T}
}

@ARTICLE{Kennicutt1998,
       author = {{Kennicutt}, Jr., Robert C.},
        title = "{The Global Schmidt Law in Star-forming Galaxies}",
      journal = {\apj},
         year = 1998,
        month = may,
       volume = {498},
       number = {2},
        pages = {541-552},
          doi = {10.1086/305588},
archivePrefix = {arXiv},
       eprint = {astro-ph/9712213},
 primaryClass = {astro-ph},
       adsurl = {https://ui.adsabs.harvard.edu/abs/1998ApJ...498..541K}
}

@ARTICLE{Rupke2011,
       author = {{Rupke}, David S.~N. and {Veilleux}, Sylvain},
        title = "{Integral Field Spectroscopy of Massive, Kiloparsec-scale Outflows in the Infrared-luminous QSO Mrk 231}",
      journal = {\apjl},
         year = 2011,
        month = mar,
       volume = {729},
       number = {2},
          eid = {L27},
        pages = {L27},
          doi = {10.1088/2041-8205/729/2/L27},
archivePrefix = {arXiv},
       eprint = {1102.4349},
 primaryClass = {astro-ph.GA},
       adsurl = {https://ui.adsabs.harvard.edu/abs/2011ApJ...729L..27R}
}

@ARTICLE{Davies2024,
       author = {{Davies}, Rebecca L. and {Belli}, Sirio and {Park}, Minjung and {Mendel}, J. Trevor and {Johnson}, Benjamin D. and {Conroy}, Charlie and {Benton}, Chlo{\"e} and {Bugiani}, Letizia and {Emami}, Razieh and {Leja}, Joel and {Li}, Yijia and {Maheson}, Gabriel and {Mathews}, Elijah P. and {Naidu}, Rohan P. and {Nelson}, Erica J. and {Tacchella}, Sandro and {Terrazas}, Bryan A. and {Weinberger}, Rainer},
        title = "{JWST reveals widespread AGN-driven neutral gas outflows in massive z   2 galaxies}",
      journal = {\mnras},
         year = 2024,
        month = mar,
       volume = {528},
       number = {3},
        pages = {4976-4992},
          doi = {10.1093/mnras/stae327},
archivePrefix = {arXiv},
       eprint = {2310.17939},
 primaryClass = {astro-ph.GA},
       adsurl = {https://ui.adsabs.harvard.edu/abs/2024MNRAS.528.4976D}
}

@ARTICLE{Cicone2014,
       author = {{Cicone}, C. and {Maiolino}, R. and {Sturm}, E. and {Graci{\'a}-Carpio}, J. and {Feruglio}, C. and {Neri}, R. and {Aalto}, S. and {Davies}, R. and {Fiore}, F. and {Fischer}, J. and {Garc{\'\i}a-Burillo}, S. and {Gonz{\'a}lez-Alfonso}, E. and {Hailey-Dunsheath}, S. and {Piconcelli}, E. and {Veilleux}, S.},
        title = "{Massive molecular outflows and evidence for AGN feedback from CO observations}",
      journal = {\aap},
         year = 2014,
        month = feb,
       volume = {562},
          eid = {A21},
        pages = {A21},
          doi = {10.1051/0004-6361/201322464},
archivePrefix = {arXiv},
       eprint = {1311.2595},
 primaryClass = {astro-ph.CO},
       adsurl = {https://ui.adsabs.harvard.edu/abs/2014A&A...562A..21C}
}

@ARTICLE{Harrison2016,
       author = {{Harrison}, C.~M. and {Alexander}, D.~M. and {Mullaney}, J.~R. and {Stott}, J.~P. and {Swinbank}, A.~M. and {Arumugam}, V. and {Bauer}, F.~E. and {Bower}, R.~G. and {Bunker}, A.~J. and {Sharples}, R.~M.},
        title = "{The KMOS AGN Survey at High redshift (KASHz): the prevalence and drivers of ionized outflows in the host galaxies of X-ray AGN}",
      journal = {\mnras},
         year = 2016,
        month = feb,
       volume = {456},
       number = {2},
        pages = {1195-1220},
          doi = {10.1093/mnras/stv2727},
archivePrefix = {arXiv},
       eprint = {1511.00008},
 primaryClass = {astro-ph.GA},
       adsurl = {https://ui.adsabs.harvard.edu/abs/2016MNRAS.456.1195H}
}

@ARTICLE{Kakkad2020,
       author = {{Kakkad}, D. and {Mainieri}, V. and {Vietri}, G. and {Carniani}, S. and {Harrison}, C.~M. and {Perna}, M. and {Scholtz}, J. and {Circosta}, C. and {Cresci}, G. and {Husemann}, B. and {Bischetti}, M. and {Feruglio}, C. and {Fiore}, F. and {Marconi}, A. and {Padovani}, P. and {Brusa}, M. and {Cicone}, C. and {Comastri}, A. and {Lanzuisi}, G. and {Mannucci}, F. and {Menci}, N. and {Netzer}, H. and {Piconcelli}, E. and {Puglisi}, A. and {Salvato}, M. and {Schramm}, M. and {Silverman}, J. and {Vignali}, C. and {Zamorani}, G. and {Zappacosta}, L.},
        title = "{SUPER. II. Spatially resolved ionised gas kinematics and scaling relations in z {\ensuremath{\sim}} 2 AGN host galaxies}",
      journal = {\aap},
         year = 2020,
        month = oct,
       volume = {642},
          eid = {A147},
        pages = {A147},
          doi = {10.1051/0004-6361/202038551},
archivePrefix = {arXiv},
       eprint = {2008.01728},
 primaryClass = {astro-ph.GA},
       adsurl = {https://ui.adsabs.harvard.edu/abs/2020A&A...642A.147K}
}

@ARTICLE{Woo2017,
       author = {{Woo}, Jong-Hak and {Son}, Donghoon and {Bae}, Hyun-Jin},
        title = "{Delayed or No Feedback? Gas Outflows in Type 2 AGNs. III.}",
      journal = {\apj},
         year = 2017,
        month = apr,
       volume = {839},
       number = {2},
          eid = {120},
        pages = {120},
          doi = {10.3847/1538-4357/aa6894},
archivePrefix = {arXiv},
       eprint = {1702.06681},
 primaryClass = {astro-ph.GA},
       adsurl = {https://ui.adsabs.harvard.edu/abs/2017ApJ...839..120W}
}

@ARTICLE{Shen2013,
       author = {{Shen}, Yue},
        title = "{The mass of quasars}",
      journal = {Bulletin of the Astronomical Society of India},
         year = 2013,
        month = mar,
       volume = {41},
       number = {1},
        pages = {61-115},
          doi = {10.48550/arXiv.1302.2643},
archivePrefix = {arXiv},
       eprint = {1302.2643},
 primaryClass = {astro-ph.CO},
       adsurl = {https://ui.adsabs.harvard.edu/abs/2013BASI...41...61S}
}

@ARTICLE{Baron2019,
       author = {{Baron}, Dalya and {M{\'e}nard}, Brice},
        title = "{Black hole mass estimation for active galactic nuclei from a new angle}",
      journal = {\mnras},
         year = 2019,
        month = aug,
       volume = {487},
       number = {3},
        pages = {3404-3418},
          doi = {10.1093/mnras/stz1546},
archivePrefix = {arXiv},
       eprint = {1903.01996},
 primaryClass = {astro-ph.GA},
       adsurl = {https://ui.adsabs.harvard.edu/abs/2019MNRAS.487.3404B}
}

@ARTICLE{Piotrowska2022,
       author = {{Piotrowska}, Joanna M. and {Bluck}, Asa F.~L. and {Maiolino}, Roberto and {Peng}, Yingjie},
        title = "{On the quenching of star formation in observed and simulated central galaxies: evidence for the role of integrated AGN feedback}",
      journal = {\mnras},
         year = 2022,
        month = may,
       volume = {512},
       number = {1},
        pages = {1052-1090},
          doi = {10.1093/mnras/stab3673},
archivePrefix = {arXiv},
       eprint = {2112.07672},
 primaryClass = {astro-ph.GA},
       adsurl = {https://ui.adsabs.harvard.edu/abs/2022MNRAS.512.1052P}
}

@ARTICLE{Baldwin1981,
       author = {{Baldwin}, J.~A. and {Phillips}, M.~M. and {Terlevich}, R.},
        title = "{Classification parameters for the emission-line spectra of extragalactic objects.}",
      journal = {\pasp},
         year = 1981,
        month = feb,
       volume = {93},
        pages = {5-19},
          doi = {10.1086/130766},
       adsurl = {https://ui.adsabs.harvard.edu/abs/1981PASP...93....5B}
}


\end{document}